# Dispersion relation for transverse waves in a linear chain of particles

V. I. Repchenkov*

**It is difficult to overestimate the importance that have for the development of science the simplest physical and mechanical models. One of them is the model of the linear one-dimensional chain of harmonic oscillators (atoms or molecules) (Fig. 1) used in the study of vibration and wave processes in a wide range of scientific disciplines, from physics of crystals, phonons to quantum chemistry and molecular biology.**

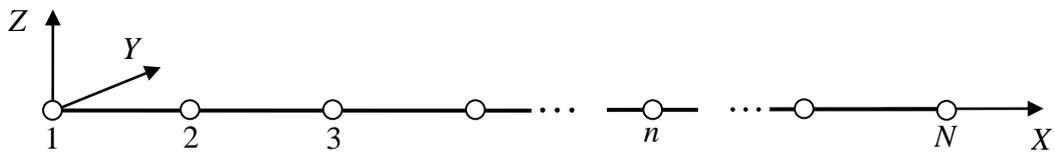

**Figure 1 − Linear chain of atoms**

A mathematical model of this system is a matrix equation of motion, which connects the acceleration of material points $\ddot{\bar{P}}$, displacement $\bar{P}$, external forces $\bar{F}$ and internal $-[k]\bar{P}$ forces:

$$[m]\ddot{\bar{P}} = \bar{F} - [k]\bar{P}. \qquad (1)$$

In the analytical mechanics and the finite element method (FEM) the matrix [m] and [k] called matrix of inertia and stiffness matrix[1,2]. In other disciplines, where the apparatus of classical mechanics is used, other terms take place. For example, [k] is called the matrix of potential energy, the force constants matrix, the Hessian of the potential energy function calculated at the equilibrium position of a system. For some reason (interdisciplinary barrier?), in physics and chemistry do not use the fundamental property of stiffness matrix formed in the Cartesian coordinate system. The assertion is valid that the columns of the matrix [k] in this representation are balanced systems of forces. More precisely, each matrix element is numerically equal to the projection of the nodal force on one of the coordinate axes, and all together the elements of a column (row) are a balanced system of forces, which is necessary to displace one of the node (mass point) per unit length along

* V. I. Repchenkov, PhD, Senior Lecturer, Belarusian State University, Minsk, Republic of Belarus.

the respective degrees of freedom, leaving all other in the rest. This follows from the equation of statics

$$[k]\overline{P} = \overline{F},\qquad(2)$$

expressing equality of the internal and external forces, when the displacement vector contains a single unit element. For example, if $\overline{P} = (1,0,0\ldots,0)$ it turns out equality for the first column:

$$[k]\overline{P} = \begin{bmatrix} k_{11} \\ k_{21} \\ \ldots \\ k_{N1} \end{bmatrix} = \begin{bmatrix} F_1 \\ F_2 \\ \ldots \\ F_N \end{bmatrix}.\qquad(3)$$

The global stiffness matrix $[k']$ of the chains of $N$ identical harmonic oscillators (atoms), performing a longitudinal movement is tridiagonal and has the form:

$$[k'] = \lambda' \begin{bmatrix} 1 & -1 & 0 & 0 & \cdots & 0 & 0 \\ -1 & 2 & -1 & 0 & \cdots & 0 & 0 \\ 0 & -1 & 2 & -1 & \cdots & 0 & 0 \\ 0 & 0 & -1 & 2 & \cdots & 0 & 0 \\ \ldots & \ldots & \ldots & \ldots & \ddots & \ldots & \ldots \\ 0 & 0 & 0 & 0 & \cdots & 2 & -1 \\ 0 & 0 & 0 & 0 & \cdots & -1 & 1 \end{bmatrix},\qquad(4)$$

this assumes the elastic interaction with constant $\lambda'$ only two nearest neighbours (valence bond). It is formed from the extended to the size $N \times N$ local stiffness matrices $[k']_e$ of elastic segments (finite elements):

$$[k']_e = \lambda' \begin{bmatrix} 1 & -1 \\ -1 & 1 \end{bmatrix}.\qquad(5)$$

Algorithm shown on (4).

As for the matrix of inertia $[m]$, it contains masses of atoms and is diagonal:

$$[m] = m \begin{bmatrix} 1 & & & \\ & 1 & & \\ & & \ddots & \\ & & & 1 \end{bmatrix},\qquad(6)$$

* V. I. Repchenkov, PhD, Senior Lecturer, Belarusian State University, Minsk, Republic of Belarus.

To the wave process in an infinite chain with interatomic distance equal to *a* corresponds the displacement field $X_n = A\exp[i(kna - \omega t)]$. By virtue of (1), (4) and (6) equations of motion have the form

$$m\ddot{X}_n = \lambda'(-X_{n-1} + 2X_n - X_{n+1}),  \quad (7)$$

where *n* is the number of the particle. It follows the classical dispersion relation between the angular frequency ω and the wave number *k*:

$$\omega = 2\sqrt{\frac{\lambda'}{m}}\sin\left(\frac{ka}{2}\right). \quad (8)$$

The application of the above stiffness matrix (4) to study the transverse movements in a linear chain, which is usually done[3,4], is incorrect. Based on the previously discussed elastic segment it is impossible to simulate the bending of the linear chain, because acting on its node (a material point), it is impossible to keep the element from turning, if to another node applied shear force. The minimum number of nodes required for that structural element could work in bending, equal to three (Fig. 2). In this case, fixing two of them can resist the transverse force acting at free node.

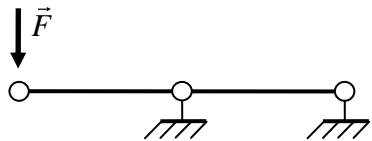

**Figure 2 − Bending element**

The stiffness matrix of bending finite element, whose nodes move only orthogonal axis *X* along the axis *Y*, build on the basis of the physical meaning of its columns. In accordance with the number of degrees of freedom of the system matrix has a dimension 3×3:

$$[k]_e = \begin{bmatrix} k_{11} & k_{12} & k_{13} \\ k_{21} & k_{22} & k_{23} \\ k_{31} & k_{32} & k_{33} \end{bmatrix} \quad (9)$$

and contains nine elements. The condition of the symmetry of the matrix $k_{ij}=k_{ji}$ reduces the number of independent force constants to six: three diagonal and three, lying under the diagonal. They, in turn, must satisfy the conditions of equilibrium of forces and moments of forces[2,5]. Displace

* V. I. Repchenkov, PhD, Senior Lecturer, Belarusian State University, Minsk, Republic of Belarus.

alternately each of the nodes on $Y_1=1$, $Y_2=1$, $Y_3=1$ leaving others in place (Fig. 3) and write the equilibrium conditions:

$$OY: \begin{aligned} k_{11}+k_{21}+k_{31}&=0,\\ k_{21}+k_{22}+k_{32}&=0,\\ k_{31}+k_{32}+k_{33}&=0, \end{aligned} \qquad M_Z: \begin{aligned} a_1\cdot k_{21}+(a_1+a_2)\cdot k_{31}&=0,\\ a_1\cdot k_{22}+(a_1+a_2)\cdot k_{32}&=0,\\ a_1\cdot k_{32}+(a_1+a_2)\cdot k_{33}&=0. \end{aligned} \qquad (10)$$

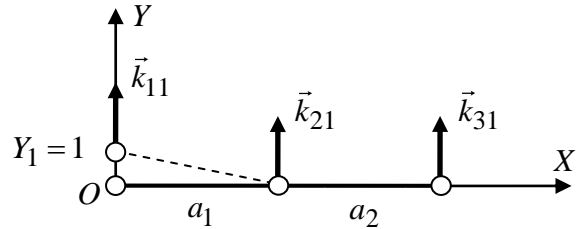

**Figure 3 − System of forces from elements of the first column of the stiffness matrix**

These six equations for the six elements of the stiffness matrix allows to express five of them in one, for example, through $k_{11}=\lambda$:

$$[k]_e = \lambda \begin{bmatrix} 1 & -\dfrac{a_1+a_2}{a_2} & \dfrac{a_1}{a_2} \\ -\dfrac{a_1+a_2}{a_2} & \left(\dfrac{a_1+a_2}{a_2}\right)^2 & -\dfrac{(a_1+a_2)a_1}{a_2^2} \\ \dfrac{a_1}{a_2} & -\dfrac{(a_1+a_2)a_1}{a_2^2} & \left(\dfrac{a_1}{a_2}\right)^2 \end{bmatrix}. \qquad (11)$$

In fact (11) corresponds to the quadratic force field of straight elastic angle between the valencies, if considered bending vibrations of linear triatomic molecules[2].

In the particular case, when $a_1=a_2=a$, the matrix (11) is simplified:

$$[k]_e = \lambda \begin{bmatrix} 1 & -2 & 1 \\ -2 & 4 & -2 \\ 1 & -2 & 1 \end{bmatrix}. \qquad (12)$$

Applying (12) to the formation of mathematical model of linear chain, which commits transverse motion, leads to the result:

* V. I. Repchenkov, PhD, Senior Lecturer, Belarusian State University, Minsk, Republic of Belarus.

$$[k] = \lambda \begin{bmatrix} 1 & -2 & 1 & 0 & 0 & \cdots & \cdots & \cdots & \cdots \\ -2 & 5 & -4 & 1 & 0 & \cdots & \cdots & \cdots & \cdots \\ 1 & -4 & 6 & -4 & 1 & \cdots & \cdots & \cdots & \cdots \\ 0 & 1 & -4 & 6 & -4 & \cdots & \cdots & \cdots & \cdots \\ 0 & 0 & 1 & -4 & 6 & \cdots & \cdots & \cdots & \cdots \\ \cdots & \cdots & \cdots & \cdots & \cdots & \ddots & \cdots & \cdots & \cdots \\ \cdots & \cdots & \cdots & \cdots & \cdots & \cdots & 6 & -4 & 1 \\ \cdots & \cdots & \cdots & \cdots & \cdots & \cdots & -4 & 5 & -2 \\ \cdots & \cdots & \cdots & \cdots & \cdots & \cdots & 1 & -2 & 1 \end{bmatrix}. \tag{13}$$

The equation of motion for *n*-th atom of an endless chain is defined by typical row of the matrix (13):

$$m\ddot{X}_n = \lambda\left(X_{n-2} - 4X_{n-1} + 6X_n - 4X_{n+1} + X_{n+2}\right). \tag{14}$$

Standard calculation now leads to another formulae in comparison with (8):

$$\omega = 4\sqrt{\frac{\lambda}{m}} \sin^2\left(\frac{ka}{2}\right) \tag{15}$$

Pay attention also to the fact that a three-node structural elements of the chain consisting of atoms of two types (Fig. 4) have different electron density distribution, and hence different rigidities. Therefore, dispersion relations $\omega_i = \omega_i(k)$, $i = 1,2$ (acoustic and optical modes) will include not only the values of the masses $M$ and $m$, but the elastic constants $\lambda_M$ and $\lambda_m$:

$$\omega_{12}^2 = \frac{2\left[M\left(\lambda_m c^2 + \lambda_M\right) + m\left(\lambda_M c^2 + \lambda_m\right)\right]}{Mm} \times$$
$$\times \left(1 \pm \sqrt{1 - \frac{4Mm\left[\left(\lambda_M c^2 + \lambda_m\right)\left(\lambda_m c^2 + \lambda_M\right) - 4\left(\lambda_M + \lambda_m\right)^2 c^2\right]}{\left[M\left(\lambda_m c^2 + \lambda_M\right) + m\left(\lambda_M c^2 + \lambda_m\right)\right]^2}}\right). \tag{16}$$

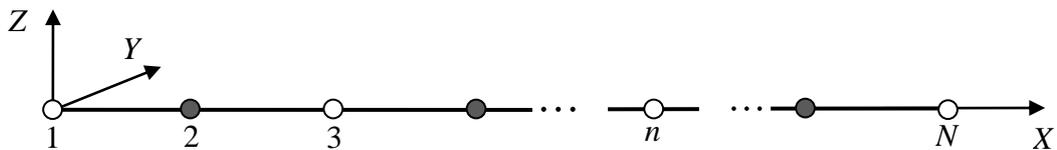

**Figure 4 – Linear chain of atoms of two types**

* V. I. Repchenkov, PhD, Senior Lecturer, Belarusian State University, Minsk, Republic of Belarus.

It is possible that the observed difference in the dispersion laws for longitudinal and transverse waves can lead to significant distinction in the results of calculations of mechanical, thermal, optical, magnetic and other properties of crystals, nanoscale and molecular systems. Also, on our view, it would be useful to improve the situation – to include this result in educational practice.

\* V. I. Repchenkov, PhD, Senior Lecturer, Belarusian State University, Minsk, Republic of Belarus.